\documentclass[twocolumn,twoside,slac_two]{revtex4}
\usepackage{graphicx}
\usepackage{fancyhdr}
\begin{document}
%%%%%%%%%%%%%%%%%%%%%%%%%%%%%%%%%%%%%%%%%%%%%%%%%
% Title page
%%%%%%%%%%%%%%%%%%%%%%%%%%%%%%%%%%%%%%%%%%%%%%%%%
%
%%%%%%%%%%%%%%%%%%%%
% Title
%%%%%%%%%%%%%%%%%%%%
\title{
Dynamical One-Armed Spiral Instability in Differentially Rotating Stars}
%
%%%%%%%%%%%%%%%%%%%%
% Author
%%%%%%%%%%%%%%%%%%%%
% Repeat the \author .. \affiliation  etc. as needed
%
% \affiliation command applies to all authors since the last
% \affiliation command. The \affiliation command should follow the
% other information

\author{Motoyuki Saijo}
\affiliation{Department of Physics, Kyoto University, Kyoto 606-8502, Japan}
\author{Shin'ichirou Yoshida}
\affiliation{
Department of Physics, University of Wisconsin-Milwaukee,
Milwaukee, WI 53211, USA}

%%%%%%%%%%%%%%%%%%%%
% Abstract
%%%%%%%%%%%%%%%%%%%%
\begin{abstract}
We investigate the dynamical one-armed spiral instability in differentially 
rotating stars with both eigenmode analysis and hydrodynamic simulations 
in Newtonian gravity.  We find that the one-armed spiral mode is generated 
around the corotation radius of the star, and the distribution of angular 
momentum shifts inwards the corotation radius during the growth of one-armed 
spiral mode.  We also find by investigating the distribution of the canonical 
angular momentum density that the low $T/|W|$ dynamical instability for both 
$m=1$ and $m=2$ mode, where $T$ is the rotational kinetic energy and $W$ is 
the gravitational potential energy, is generated around the corotation point.
Finally, we discuss the feature of gravitational waves generated from these 
modes.
\end{abstract}

%\maketitle must follow title, authors, abstract
\maketitle

\thispagestyle{fancy}

% body of paper here - Use proper section commands
% References should be done using the \cite, \ref, and \label commands
% Put \label in argument of \section for cross-referencing
%\section{\label{}}

%%%%%%%%%%%%%%%%%%%%%%%%%%%%%%%%%%%%%%%%%%%%%%%%%
%%%%%%%%%%%%%%%%%%%%%%%%%%%%%%%%%%%%%%%%%%%%%%%%%
%%%
%%%   Section I : Introduction
%%%
%%%%%%%%%%%%%%%%%%%%%%%%%%%%%%%%%%%%%%%%%%%%%%%%%
%%%%%%%%%%%%%%%%%%%%%%%%%%%%%%%%%%%%%%%%%%%%%%%%%
%%%%%%%%%%%%%%%%%%%%
\section{Introduction}
%%%%%%%%%%%%%%%%%%%%
Stars in nature are usually rotating and may be subject to non-axisymmetric 
rotational instabilities.  An exact treatment of these instabilities exists 
only for incompressible equilibrium fluids in Newtonian gravity, e.g. 
\cite{Chandra69,Tassoul}.  For these configurations, global rotational 
instabilities may arise from non-radial toroidal modes $e^{im\varphi}$ (where 
$m=\pm 1,\pm 2, \dots$ and $\varphi$ is the azimuthal angle).

For sufficiently rapid rotation, the $m=2$ bar mode becomes either 
{\em secularly} or {\em dynamically} unstable.  The onset of instability can 
typically be marked by a critical value of the dimensionless parameter 
$\beta \equiv T/|W|$, where $T$ is the rotational kinetic energy and $W$ the 
gravitational potential energy.  Uniformly rotating, incompressible stars in 
Newtonian theory are secularly unstable to bar-mode formation when $\beta 
\geq \beta_{\rm sec} \simeq 0.14$.  This instability can grow only in the 
presence of some dissipative mechanism, like viscosity or gravitational 
radiation, and the associated growth timescale is the dissipative timescale,
which is usually much longer than the dynamical timescale of the system.  By 
contrast, a dynamical instability to bar-mode formation sets in when $\beta 
\geq \beta_{\rm dyn} \simeq 0.27$.  This instability is independent of any 
dissipative mechanism, and the growth time is the hydrodynamic timescale.

There are several papers indicating that dynamical instability of the rotating 
stars occurs at low $T/|W|$, which is far below the standard criterion of 
dynamical instability in Newtonian gravity.  Tohline and Hachisu \cite{TH90} 
find in the self-gravitating ring and tori that $m=2$ dynamical instability 
occurs around $T/|W| \sim 0.16$ in the lowest case in the centrally condensed 
configurations.  Shibata, Karino, and Eriguchi \cite{SKE} find that $m=2$ 
dynamical instability occurs around $T/|W| \sim O(10^{-2})$ in the high degree 
($\Omega_{\rm c} / \Omega_{\rm eq} \approx 10$) of differential rotation.  
Note that $\Omega_{\rm c}$ and $\Omega_{\rm eq}$ are the angular velocity at 
the center and at the equatorial surface, respectively.  Centrella et al. 
\cite{CNLB01} found dynamical $m=1$ instability around $T/|W| \sim 0.09$ in 
the $n=3.33$ polytropic toroidal star with high degree ($\Omega_{\rm c} / 
\Omega_{\rm eq} = 26$) of differential rotation, and Saijo, Baumgarte, and 
Shapiro \cite{SBS03} extended the results of dynamical $m=1$ instability to 
$n \gtrsim 2.5$, $\Omega_{\rm c} / \Omega_{\rm eq} \gtrsim 10$.  Note that $n$ 
is the polytropic index of the star.

There are some indications that corotation resonance triggers dynamical bar 
instability.  Papaloizou and Pringle \cite{PP84} found that 
non-selfgravitating tori with constant specific angular momentum are
unstable to low order non-axisymmetric modes and that the modes grow on a 
dynamical time-scale.  Watts, Andersson, and Jones \cite{WAJ05} argue that the 
shear instabilities occur when the degree of differential rotation exceeds a 
critical value and the $f$-mode develops a corotation point associated with 
the presence of a continuous spectrum.  They also point out that dynamical bar 
instability that Shibata et al. \cite{SKE} found is in the corotation band.

%%%%%%%%%%%%%%%%%%%%%%%%%%%%%%%%%%%%%%%%%%%%%%%%%
%   Table 1
%%%%%%%%%%%%%%%%%%%%%%%%%%%%%%%%%%%%%%%%%%%%%%%%%
%%%%%%%%%%
\begin{table*}
\begin{center}
\begin{minipage}{14cm}
\caption{
Three differentially rotating equilibrium stars that trigger dynamical 
instability
\label{tab:initial}}
%\begin{ruledtabular}
\begin{tabular}{c c c c c c c c c}
\hline
\hline
Model &
$n$ \footnotemark[1] &
$d / R_{\rm eq}$ \footnotemark[2] &
$R_{\rm p} / R_{\rm eq}$ \footnotemark[3] &
$\Omega_{\rm c} / \Omega_{\rm eq}$ \footnotemark[4] & 
$\rho_{\rm c} / \rho_{\rm max}$ \footnotemark[5] &
$R_{\rm maxd}/R_{\rm eq}$ \footnotemark[6] & 
$T/|W|$ \footnotemark[7] & dominant unstable mode
\\
\hline
I & $3.33$ &
$0.20$ & $0.413$ & $26.0$ & $0.491$ & $0.198$ &
$0.146$ & $m=1$
\\
II & $1.00$ &
$0.20$ & $0.250$ & $26.0$ & $0.160$ & $0.383$ &
$0.119$ & $m=2$
\\
III & $1.00$ &
$1.00$ & $0.250$ & $2.0$ & $0.837$ & $0.388$ &
$0.277$ & $m=2$
\\
\hline
\hline
\end{tabular}
\footnotetext[1]{$n$: Polytropic index}
\footnotetext[2]{$R_{\rm eq}$: Equatorial radius}
\footnotetext[3]{$R_{\rm pl}$: Polar radius}
\footnotetext[4]{$\Omega_{\rm c}$: 
Central angular velocity; $\Omega_{\rm eq}$: Equatorial angular velocity}
\footnotetext[5]{$\rho_{\rm c}$: Central density; 
$\rho_{\rm max}$: Maximum density}
\footnotetext[6]{$R_{\rm maxd}$: Radius of 
maximum density}
\footnotetext[7]{$T$: Rotational kinetic energy;
$W$: Gravitational potential energy}
\end{minipage}
\end{center}
\end{table*}
%%%%%%%%%%

Our purpose in this paper is to investigate the nature of dynamical $m=1$ 
instability with both eigenmode analysis and hydrodynamical analysis.  
A non-linear hydrodynamical simulation is indispensable for investigation
of evolutionary process and final outcome of instability. The nature of
instability as a source of gravitational wave, which interests us most, is
only accessible through non-linear hydrodynamical computations. On the
other hand, a linear eigenmode analysis is in general easier to approach
the dynamical instability of given equilibria and it may be helpful
to have physical insight on the mechanism and the origin of instability.
Therefore a linear eigenmode analysis and a non-linear simulation are 
complementary to each other and they both help us to understand the
nature of dynamical instability.

For a hydrodynamical simulation, we used the numerical code developed in
Ref. \cite{SBS03}, while we introduced a toy cylinder model that mimics
differentially rotating stars to study the instability.  Self-gravitating 
cylinder models have been used to study general dynamical nature of such 
gaseous masses as stars, accretion disks and of stellar system as galaxies.  
Although there is no cylinder with infinite length in reality, it is expected 
to share some qualitative similarities with realistic astrophysical objects, 
e.g. \cite{Ostriker65}.  In fact, it has served as a useful model to 
investigate secular and dynamical instabilities of rotating masses.

This paper is organized as follows.  In \S~\ref{sec:Nhydro} we present our 
hydrodynamical results of dynamical one-armed spiral and dynamical bar 
instability.  We present our diagnosis of dynamical $m=1$ and $m=2$ 
instability by using canonical angular momentum in \S~\ref{sec:Canonical}, 
and briefly summarize our findings in \S~\ref{sec:Discussion}.  Throughout 
this paper we use gravitational units with $G = 1$.  A more detailed 
discussion will be presented in Ref. \cite{SY05}.

%%%%%%%%%%%%%%%%%%%%%%%%%%%%%%%%%%%%%%%%%%%%%%%%%
%%%%%%%%%%%%%%%%%%%%%%%%%%%%%%%%%%%%%%%%%%%%%%%%%
%%%
%%%   Section II : Dynamical instabilities in differentially rotating stars
%%%
%%%%%%%%%%%%%%%%%%%%%%%%%%%%%%%%%%%%%%%%%%%%%%%%%
%%%%%%%%%%%%%%%%%%%%%%%%%%%%%%%%%%%%%%%%%%%%%%%%%
%%%%%%%%%%%%%%%%%%%%
\section{Dynamical instabilities in differentially rotating stars}
\label{sec:Nhydro}
%%%%%%%%%%%%%%%%%%%%
First we explain features of our initial data sets of differentially 
rotating stars on which we performed non-linear hydrodynamical computations.
We assume a polytropic equation of state,
%%%%%%%%%%
\begin{equation}
P = \kappa \rho^{1+1/n},
\end{equation}
%%%%%%%%%%
where $P$ is the pressure, $\kappa$ is a constant, $\rho$ is the density, 
$n$ is the polytropic index.  One feature of the polytropic equation of state 
is that all matter quantities can be renormalized in terms of $\kappa$ so that 
$\kappa$ does not explicitly appear.  We also assume the ``$j$-constant'' 
rotation law as
%%%%%%%%%%
\begin{equation}
\Omega = \frac{j_{0}}{d^{2} + \varpi^{2}},
\label{eqn:omega}
\end{equation}
%%%%%%%%%%
where $\Omega$ is the angular velocity, $j_{0}$ is a constant parameter with 
units of specific angular momentum, and $\varpi$ is the cylindrical radius.  
The parameter $d$ determines the length scale over which $\Omega$ changes; 
uniform rotation is achieved in the limit $d \rightarrow \infty$, with keeping 
the ratio $j_0/d^2$ finite.  We choose the same data sets as Ref. \cite{SBS03} 
for investigating low $T/|W|$ dynamical instability in differentially rotating 
stars.  We also construct the equilibrium of a star with weak differential 
rotation in high $T/|W|$, which excites the standard dynamical bar 
instability, e.g. \cite{Chandra69}.  The characteristic parameters are 
summarized in Table \ref{tab:initial}.

To enhance any $m=1$ or $m=2$ instability, we disturb the initial equilibrium 
density $\rho_{\rm eq}$ by a non-axisymmetric perturbation according to
%%%%%%%%%%
\begin{equation}
\rho = \rho_{\rm eq}
\left( 1 +
  \delta^{(1)} \frac{x+y}{R_{\rm eq}} +
  \delta^{(2)} \frac{x^{2}-y^{2}}{R_{\rm eq}^{2}}
\right),
\label{eqn:DPerturb}
\end{equation}
%%%%%%%%%%
where $R_{\rm eq}$ is the equatorial radius, with 
$\delta^{(1)} = \delta^{(2)} \approx 1.7 - 2.8 \times 10^{-3}$ in all our 
simulations.  We also introduce ``dipole'' $D$ and ``quadrupole'' $Q$ 
diagnostics to monitor the development of $m=1$ and $m=2$ modes as
%%%%%%%%%%
\begin{eqnarray}
D &=& \left< e^{i m \varphi} \right>_{m=1} =
\frac{1}{M} \int \rho \frac{x + i y}{\sqrt{x^{2}+y^{2}}} d^3 x,
\\
Q &=& \left< e^{i m \varphi} \right>_{m=2} =
\frac{1}{M} \int \rho \frac{(x^{2}-y^{2}) + i (2 x y)}{x^{2}+y^{2}} d^3 x,
\nonumber \\
\end{eqnarray}
%%%%%%%%%%
respectively.

We compute approximate gravitational waveforms by using the quadrupole 
formula.  In the radiation zone, gravitational waves can be described by a 
transverse-traceless, perturbed metric $h_{ij}^{TT}$ with respect to a flat 
spacetime. In the quadrupole formula, $h_{ij}^{TT}$ is found from 
%%%%%%%%%%
\begin{equation}
h_{ij}^{TT}= \frac{2}{r} \frac{d^{2}}{d t^{2}} I_{ij}^{TT},
\label{eqn:wave1}
\end{equation}
%%%%%%%%%%
where $r$ is the distance to the source, $I_{ij}$ the quadrupole moment of 
the mass distribution, and where $TT$ denotes the transverse-traceless 
projection.  Choosing the direction of the wave propagation to be along the 
rotational axis ($z$-axis), the two polarization modes of gravitational 
waves can be determined from
%%%%%%%%%%
\begin{equation}
h_{+} \equiv \frac{1}{2} (h_{xx}^{TT} - h_{yy}^{TT})
\mbox{~~~and~~~}
h_{\times} \equiv h_{xy}^{TT}.
\end{equation}
%%%%%%%%%%
For observers along the rotation axis, we thus have
%%%%%%%%%%
\begin{eqnarray}
\frac{r h_{+}}{M} &=&
\frac{1}{2 M} \frac{d^{2}}{d t^{2}} ( I_{xx}^{TT} - I_{yy}^{TT}), \label{h+}
\\
\frac{r h_{\times}}{M} &=&
\frac{1}{M} \frac{d^{2}}{d t^{2}} I_{xy}^{TT} \label{h-}
.
\end{eqnarray}
%%%%%%%%%%

%%%%%%%%%%%%%%%%%%%%%%%%%%%%%%%%%%%%%%%%%%%%%%%%%%%
%%% Figure 1
%%%%%%%%%%%%%%%%%%%%%%%%%%%%%%%%%%%%%%%%%%%%%%%%%%%
%%%%%%%%%%%%
\begin{figure}
\centering
\includegraphics[width=7cm]{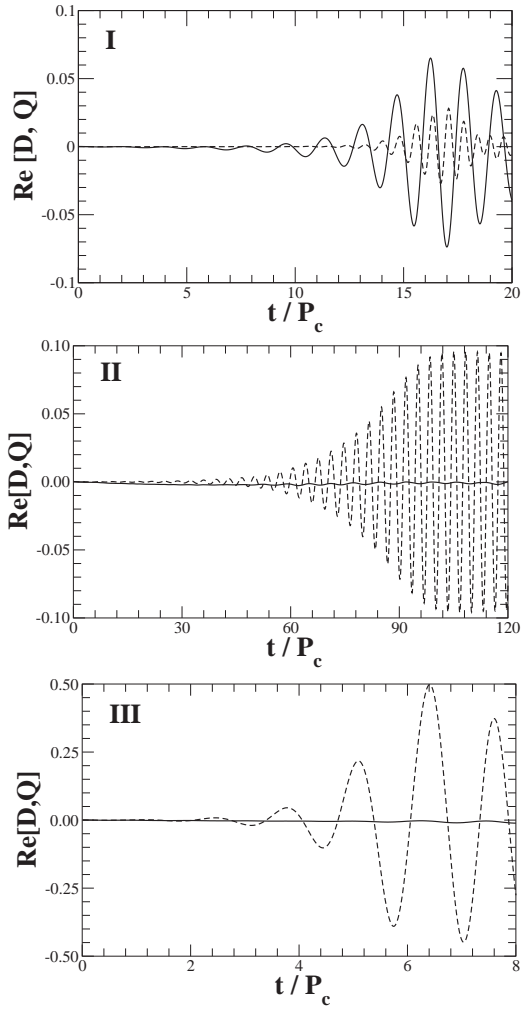}
\caption{
Diagnostics $D$ and $Q$ as a function of $t/P_{\rm c}$ for 3 differentially 
rotating stars (see Table \ref{tab:initial}).  Solid and dotted lines
denote the values of $D$ and $Q$, respectively.  The Roman numeral
in each panel corresponds to the model of the differentially rotating
stars, respectively.  Hereafter $P_{\rm c}$ represents the central rotation 
period.
}
\label{fig:dig}
\end{figure}
%%%%%%%%%%%%

%%%%%%%%%%%%%%%%%%%%%%%%%%%%%%%%%%%%%%%%%%%%%%%%%%%
%%% Figure 2
%%%%%%%%%%%%%%%%%%%%%%%%%%%%%%%%%%%%%%%%%%%%%%%%%%%
%%%%%%%%%%%%
\begin{figure}
\centering
\includegraphics[width=7.0cm]{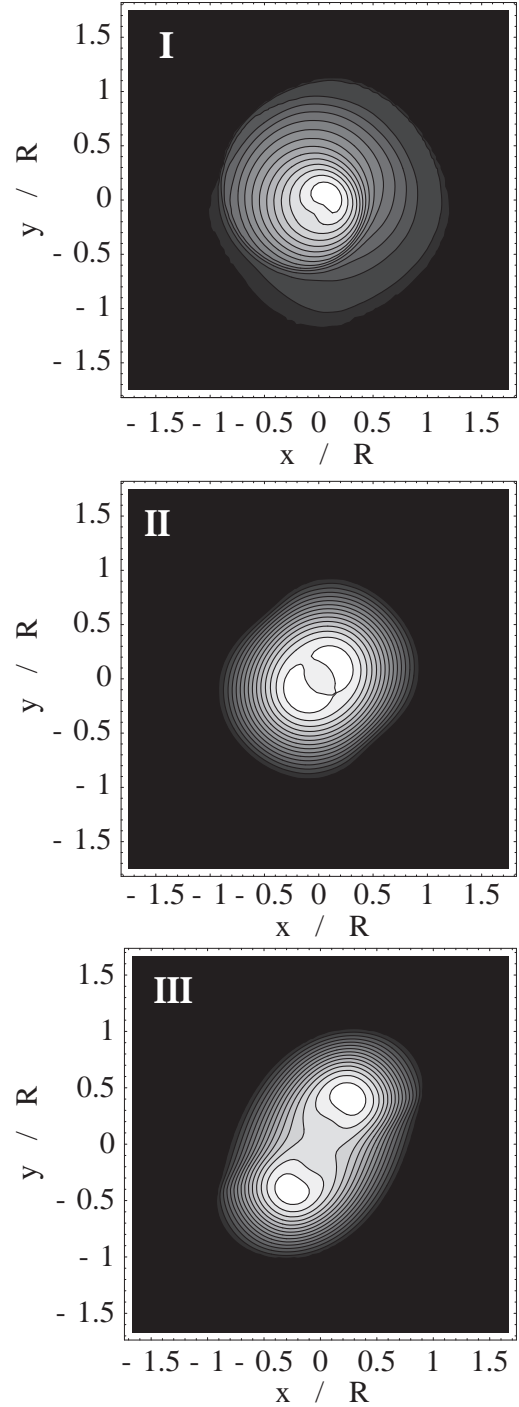}
\caption{
Density contours in the equatorial plane for 3 differentially rotating stars
(see Table \ref{tab:initial}).  
Models I, II, and III are plotted at the parameter ($t/P_{\rm c}$, 
$\rho_{\rm max} / \rho_{\rm max}^{(0)}$) =
($16.3$, $3.63$), 
($132$, $1.25$), and 
($5.49$, $1.20$), 
where $\rho_{\rm max}$ is the maximum density, $\rho_{\rm max}^{(0)}$ is the 
maximum density at $t=0$, and $R$ is the equatorial radius at $t=0$.  The 
contour lines denote densities $\rho/\rho_{\rm max} = 10^{- (16-i) d}
(i=1, \cdots, 15)$ for model I and $\rho/\rho_{\rm max} = 6.67 (16-i) \times
10^{-2} (i=1, \cdots, 15)$ for models II and III, respectively.
}
\label{fig:qxy}
\end{figure}
%%%%%%%%%%%%

The time evolutions of dipole diagnostic and the quadrupole diagnostic are 
plotted in Fig. \ref{fig:dig}.  We determine that the system is stable 
to $m=1$ ($m=2$) mode when the dipole (quadrupole) diagnostic remains small 
throughout the evolution, while the system is unstable when the diagnostic 
grows exponentially at the early stage of the evolution.  It is clearly seen 
in Fig. \ref{fig:dig} that the star is more unstable to the one-armed spiral 
mode for model I, and more unstable to the bar mode for models II and III.  
In fact, both diagnostics grow for model I.  The dipole diagnostic, however, 
grows more markedly than the quadrupole diagnostic, indicating that the $m=1$ 
mode is the dominant unstable mode.

The density contour of the differentially rotating stars are shown in Fig. 
\ref{fig:qxy}.  It is clearly seen in Fig. \ref{fig:qxy} that one-armed
spiral structure is formed at the intermediate stage of the evolution
for model I, and that bar structure is formed for models II and III once the 
dynamical instability sets in.

\begin{figure}
\centering
\includegraphics[width=7cm]{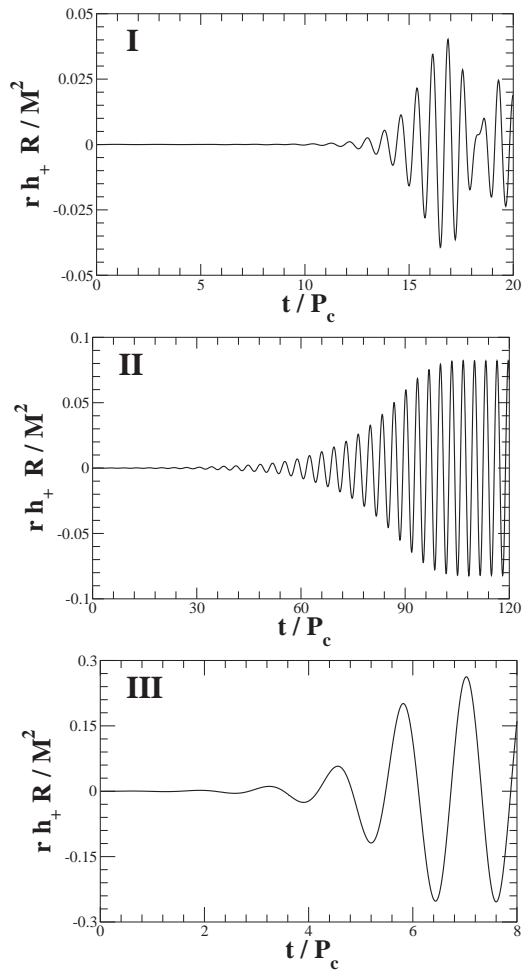}
\caption{
Gravitational waveform for 3 differentially rotating stars (see Table 
\ref{tab:initial}) as seen by a distant observer located on the rotational 
axis of the equilibrium star.
}
\label{fig:gw}
\end{figure}

We also show gravitational waves generated from dynamical one-armed spiral and 
bar instability in Fig. \ref{fig:gw}.  For $m=1$ modes, the gravitational
radiation is emitted not by the primary mode itself, but by the $m=2$
secondary harmonic which is simultaneously excited, albeit at a lower
amplitude.  Unlike the case for bar-unstable stars, the gravitational wave 
signal does not persist for many periods, but instead is damped fairly rapidly.

%%%%%%%%%%%%%%%%%%%%%%%%%%%%%%%%%%%%%%%%%%%%%%%%%
%%%%%%%%%%%%%%%%%%%%%%%%%%%%%%%%%%%%%%%%%%%%%%%%%
%%%
%%%   Section IV : Canonical Angular Momentum
%%%
%%%%%%%%%%%%%%%%%%%%%%%%%%%%%%%%%%%%%%%%%%%%%%%%%
%%%%%%%%%%%%%%%%%%%%%%%%%%%%%%%%%%%%%%%%%%%%%%%%%
%%%%%%%%%%
\section{Canonical Angular Momentum}
\label{sec:Canonical}
%%%%%%%%%%
We introduce canonical angular momentum following Ref. \cite{FS78} to diagnose 
oscillations in rotating fluids.  For adiabatic linear perturbations on a 
perfect fluid configuration in stationary, axisymmetric spacetime, it is 
possible to introduce canonical conserved quantities.  Since we only use 
canonical angular momentum $J_{\rm c}$ in this paper, we write down its 
explicit form as
%%%%%%%%%%
\begin{eqnarray}
J_{\rm c} &=& 
m\int (\Re[\sigma]-m\Omega)\rho|\xi|^2 d^{3}x
\nonumber \\
&&
-2m\int \rho \varpi\Omega\cdot \Im[\xi^\varpi\xi^{\varphi *}] d^{3}x,
\label{canonJform}
\end{eqnarray}
%%%%%%%%%%
where $\sigma$ is the eigenfrequency, $\xi^{i}$ is Lagrangian  displacement 
vector.  Note that total canonical angular momentum becomes zero when 
dynamical instability sets in.

%%%%%%%%%%%%%%%%%%%%%%%%%%%%%%%%%%%%%%%%%%%%%%%%%%%
%%% Figure 3
%%%%%%%%%%%%%%%%%%%%%%%%%%%%%%%%%%%%%%%%%%%%%%%%%%%
%%%%%%%%%%%%
%%%%%%%%%%%%

\begin{figure*}
\centering
\includegraphics[width=17cm]{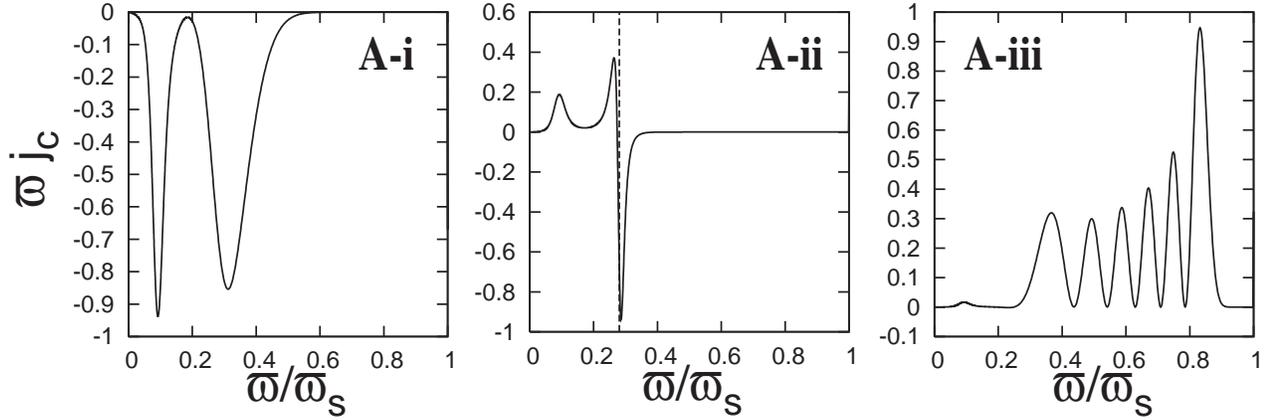}
\caption{
Distribution of canonical angular momentum density for $m=1$ unstable mode 
(see Table \ref{tab:cylfrq}).  Vertical dashed line represents the location 
of corotation radius of the mode.  
The Roman character in each panel corresponds to the model of the cylindrical 
gaseous fluid, respectively.
Note that we normalized the distribution 
of the canonical angular momentum in an appropriate value, since the 
eigenfunction can be scaled arbitrarily.
}
\label{fig:jcm1}
\end{figure*}

\begin{figure}
\centering
\includegraphics[width=7cm]{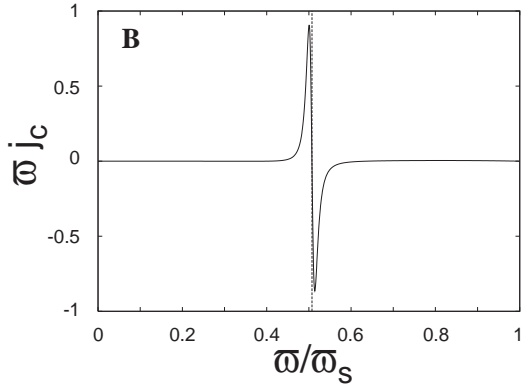}
\caption{
Distribution of canonical angular momentum density for $m=2$ unstable mode 
(see Table \ref{tab:cylfrq}). 
The Roman character in the panel corresponds to the model of the cylindrical 
gaseous fluid.
Vertical dashed lines mark the locations of corotation radius of the mode.
}
\label{fig:jcm2}
\end{figure}

Next we apply the method of canonical angular momentum to the linearized 
oscillations of a cylinder.  We prepare two $m=1$ stable modes (A-i, A-iii) 
and one $m=1$ unstable mode (A-ii), summarized in Table \ref{tab:cylfrq}.  We 
plot the integrand of canonical angular momentum $\varpi j_{\rm c}$ 
%%%%%
\begin{equation}
j_{c} = m (\Re[\sigma]-m\Omega)\rho|\xi|^2
   -2m \rho \varpi\Omega\cdot \Im[\xi^\varpi\xi^{\varphi *}],
\end{equation}
%%%%%
for $m=1$ mode in Fig. \ref{fig:jcm1}.  We define corotation radius 
$\varpi_{\rm crt}$ of modes as $\Re[\sigma]-m\Omega(\varpi)=0$.  This means 
that the pattern speed of mode coincides with the local rotational frequency 
of background flow there.  We find that the distribution of canonical angular 
momentum changes its sign around corotation radius in $m=1$ unstable case, and 
that it is positive for $\varpi<\varpi_{\rm{crt}}$, while it is negative for 
$\varpi>\varpi_{\rm{crt}}$.  The behavior of the canonical angular momentum 
suggests us that this instability is related to the existence of corotation 
point inside the cylinder.  We also find that the positive case of the 
canonical angular momentum (A-iii) corresponds to the case when the pattern 
speed of the mode is faster than the rotation of cylinder in all radius, while 
the negative (A-i) corresponds to the opposite.  We also check the low $T/|W|$ 
bar instability in cylindrical model and find that the same behavior as $m=1$ 
mode (B) appears in the distribution of the canonical angular momentum density 
(Fig. \ref{fig:jcm2}).

%%%%%%%%%%%%%%%%%%%%%%%%%%%%%%%%%%%%%%%%%%%%%%%%%%%
%%% Table 2
%%%%%%%%%%%%%%%%%%%%%%%%%%%%%%%%%%%%%%%%%%%%%%%%%%%
%%%%%%%%%%
\begin{table}
\begin{center}
\caption{Parameters for equilibrium gaseous fluid and eigenfrequency 
\label{tab:cylfrq}
}
\begin{tabular}{c c c c c}
\hline
\hline
Model  &  
$\Omega_{\rm c}/\Omega_{\rm s}$  \footnotemark[8] & 
$T/|W|$ & $\sigma/\Omega_{\rm c}$   \footnotemark[9] &
$\varpi_{\rm{crt}}/\varpi_{\rm s}$ \footnotemark[10]
\\
\hline
A-i   & 11.34 & 0.460 & $-0.245$ & ---\\
A-ii  & 11.34 & 0.460 & $0.551+0.0315 i$ & 0.281\\
A-iii & 11.34 & 0.460 & $1.15$ & ---\\
B & 13.00 & 0.170 & $0.327+0.0126 i$ & 0.507\\
\hline
\hline
\end{tabular}
\footnotetext[8]{$\Omega_{\rm s}$: Surface angular velocity}
\footnotetext[9]{$\sigma$: Eigenfrequency}
\footnotetext[10]{$\varpi_{\rm crt}$: Corotation radius; 
$\varpi_{\rm s}$: Surface radius}
\end{center}
\end{table}
%%%%%%%%%%

%%%%%%%%%%%%%%%%%%%%%%%%%%%%%%%%%%%%%%%%%%%%%%%%%
% Figure 4
%%%%%%%%%%%%%%%%%%%%%%%%%%%%%%%%%%%%%%%%%%%%%%%%%
%%%%%%%%%%
%%%%%%%%%%%%

%%%%%%%%%%%%%%%%%%%%%%%%%%%%%%%%%%%%%%%%%%%%%%%%%
% Figure 5
%%%%%%%%%%%%%%%%%%%%%%%%%%%%%%%%%%%%%%%%%%%%%%%%%
%%%%%%%%%%
%%%%%%%%%%%%

We furthermore investigate $m=2$ instability of Bardeen disk and classical bar 
instability of Maclaurin spheroid.  In these cases, the canonical angular 
momentum density that is analytically obtained is zero in all cylindrical 
radius.  Therefore the behavior of the canonical angular momentum density 
shows a clear contrast between $m=1$, $2$ instability with high degree of 
differential rotation and $m=2$ classical instability of uniformly rotating 
fluid.

Finally, we adopt the method of canonical angular momentum to the nonlinear 
hydrodynamics.  We identify the complex eigenfrequency and the corotation 
radius from dipole or quadrupole diagnostics which is summarized in 
Table \ref{tab:freq}.  Note that we read off the eigenfrequency from
those plots at the early stage of the evolution.  The Eulerian perturbed 
velocity is defined by subtracting the velocity at equilibrium from the 
velocity.  The Lagrangian displacement vector is extracted by using a linear 
formula for a dominant mode in each case.

%%%%%%%%%%%%%%%%%%%%%%%%%%%%%%%%%%%%%%%%%%%%%%%%%%%
%%% Table 3
%%%%%%%%%%%%%%%%%%%%%%%%%%%%%%%%%%%%%%%%%%%%%%%%%%%
%%%%%%%%%%
\begin{table}
\begin{center}
\caption{
Eigenfrequency and the corotation radius of 3 differentially rotating stars
\label{tab:freq}}
%\begin{ruledtabular}
\begin{tabular}{c c c}
\hline
\hline
Model &
$\sigma$ $[\Omega_{\rm c}]$ &
$\varpi_{\rm crt}$  $[R_{\rm eq}]$
\\
\hline
I & $0.590 + 0.0896 i$ & $0.167$
\\
II & $0.284 + 0.0121 i$ & $0.492$ 
\\
III & $0.757 + 0.200 i$ & ---
\\
\hline
\hline
\end{tabular}
\end{center}
\end{table}
%%%%%%%%%%

\begin{figure}
\centering
\includegraphics[width=7cm]{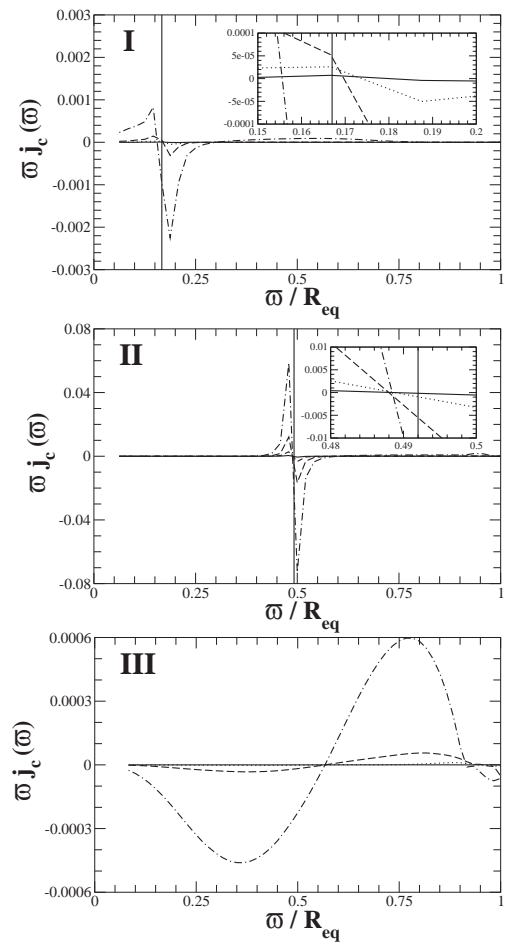}
\caption{
Snapshots of the canonical angular momentum distribution
$\varpi j_{\rm c}(\varpi)$ in the equatorial plane for 3 differentially 
rotating stars (see Table \ref{tab:initial}).  Solid, dotted, dashed, and 
dash-dotted line represents the time
$t/P_{\rm c} =$(3.47, 6.93, 10.40, 13.86) for model I,
$t/P_{\rm c} =$(45.68, 56.43, 67.18, 77.97) for model II, and 
$t/P_{\rm c} =$(1.10, 2.19, 3.29, 4.39) for model III, respectively.
Note that vertical line in panels I and II denotes the corotation radius of 
the star (model III does not have a corotation radius).  We also enlarged the 
figure around the corotation radius for panels I and II, which is presented in 
the right upper part of each panel.  Although our method of determining the 
corotation radius is not precise, we clearly find that the distribution 
significantly changes around the corotation radius.
}
\label{fig:jc}
\end{figure}

We show the snapshots of canonical angular momentum density in 
Fig. \ref{fig:jc}.  Since we determine the corotation radius using the 
extracted eigenfrequency and the angular velocity profile at equilibrium, the 
radius does not change throughout the evolution.  For low $T/|W|$ dynamical 
instability, the distribution of the canonical angular momentum drastically 
changes its sign around the corotation radius, and the maximum amount of 
canonical angular momentum density increases at the early stage of evolution.  
This means that the angular momentum flows inside the corotation radius in 
the evolution.  On the other hand, for high $T/|W|$ dynamical instability
that is related to the classical bar instability, the distribution of the 
canonical angular momentum is smooth with no particular feature and tends to 
have a positive portion outside.  This means that the canonical angular 
momentum flows outwards in the evolution, which is in clear contrast to the 
case of low $T/|W|$ one.  

From these different behaviors of the distribution of the canonical angular 
momentum, we see that the mechanisms working in the low $T/|W|$ instabilities 
and the classical bar instability may be quite different, i.e., in the former 
the corotation resonance of modes are essential, while the instability is 
global in the latter case.

%%%%%%%%%%%%%%%%%%%%%%%%%%%%%%%%%%%%%%%%%%%%%%%%%
%%%%%%%%%%%%%%%%%%%%%%%%%%%%%%%%%%%%%%%%%%%%%%%%%
%%%
%%%   Section IV : Discussion
%%%
%%%%%%%%%%%%%%%%%%%%%%%%%%%%%%%%%%%%%%%%%%%%%%%%%
%%%%%%%%%%%%%%%%%%%%%%%%%%%%%%%%%%%%%%%%%%%%%%%%%
%%%%%%%%%%
\section{Discussion}
\label{sec:Discussion}
%%%%%%%%%%

We have studied the nature of dynamical one-armed spiral instability in 
differentially rotating stars both in linear eigenmode analysis and 
in hydrodynamic simulation using canonical angular momentum distribution.

We have found that the one-armed spiral instability occurs around the 
corotation radius of the star by investigating the distribution of the 
canonical angular momentum.  We have also found by investigating the canonical 
angular momentum that the instability grows through the inflow of the angular 
momentum inside the corotation radius.  The feature also holds for the 
dynamical bar instability in low $T/|W|$, which is in clear contrast to that 
of classical dynamical bar instability in high $T/|W|$.  
Therefore the existence of 
corotation point inside the star plays a significant role of exciting 
one-armed spiral mode and bar mode dynamically in low $T/|W|$.

%%%%%%%%%%%%%%%%%%%%%%%%%%%%%%%%%%%%%%%%%%%%%%%%%
% Figure 6
%%%%%%%%%%%%%%%%%%%%%%%%%%%%%%%%%%%%%%%%%%%%%%%%%
%%%%%%%%%%
%%%%%%%%%%%%

Finally, we mention the feature of gravitational waves generated from this 
instability.  Quasi-periodic gravitational waves emitted by stars with $m=1$
instabilities have smaller amplitudes than those emitted by stars unstable to 
the $m=2$ bar mode.  For $m=1$ modes, the gravitational radiation is emitted 
not by the primary mode itself, but by the $m=2$ secondary harmonic which is 
simultaneously excited, possibly through non-linear self-coupling of m=1 mode.
(Remarkably the precedent studies \cite{CNLB01,SBS03} found that the pattern 
speed of $m=2$ mode is almost the same as that of $m=1$ mode, which suggest 
the former is the harmonic of the latter.)  Unlike the case for bar-unstable 
stars, the gravitational wave signal does not persist for many periods, but 
instead is damped fairly rapidly.

\end{document}